\documentclass[aps,prc,preprint,groupedaddress,showkeys,showpacs]{revtex4-1}
\usepackage{graphicx}% Include figure files

\usepackage{dcolumn}% Align table columns on decimal point

\usepackage{bm}%

\usepackage[dvipsnames]{xcolor}

\usepackage{epstopdf}

\usepackage{amsmath}

\usepackage[caption = false]{subfig}
\begin{document}

%\preprint{APS/123-QED}

\title{General relativistic calculation of magnetic field and Power loss for a misaligned pulsar}% Force line breaks with \\

\author{Sagnik Chatterjee}

 \email{sagnik18@iiserb.ac.in}

 \affiliation{Department of Physics, Indian Institute of Science Education and Research Bhopal.}

\author{Ritam Mallick}

 \email{mallick@iiserb.ac.in}

\affiliation{Department of Physics, Indian Institute of Science Education and Research Bhopal.}

\author{Debojoti Kuzur}

 \email{debojoti16@iiserb.ac.in}

\affiliation{Department of Physics, Indian Institute of Science Education and Research Bhopal.}

\date{\today}% It is always \today, today,


\begin{abstract}

In this study, we model a pulsar as a general relativistic oblique rotator, where the oblique rotator is a rotationally deformed neutron star whose rotation and magnetic axis are inclined at an angle. The oblique rotator spins down, losing rotational energy through the magnetic poles. The magnetic field is assumed to be dipolar; however, the star has a non-zero azimuthal component due to the misalignment. The magnetic field induces an electric field for a force-free condition. The magnetic field decreases as the misalignment increases and is minimum along the equatorial plane of the star.
In contrast, the electric field remains almost constant initially but decreases rapidly at a high misalignment angle. The charge separation at the star surface is qualitatively similar to that of Newtonian calculation. We find that the power loss for a general relativistic rotator is minimum for either an aligned or an orthogonal rotator, which contrasts with Newtonian calculation, where the power loss increases with an increase in the misalignment angle.

\end{abstract}



\maketitle

\section{Introduction}

Observed pulsars are considered as rotating neutron stars (NS) having a dipole magnetic field and rotating at an axis that is different from the magnetic axis \cite{gold,pacini}. This scenario is also known as an oblique rotator. Neutron stars are formed after the supernova explosion, and assuming magnetic flux and angular momentum conservation, they are born with a high magnetic field and rotation rate. As the rotation and magnetic axis (from where the emission occurs) do not coincide, the rotating neutron star emits very regular pulsations and are termed pulsars. Therefore, it behaves as an inclined rotating dipole that is emitting radiating and losing rotational energy \cite{Kuzur,Cerutti}. The study of pulsar or neutron stars has gained recent interest due to their unique feature, which is still not realizable on earth based experiments. NS are very dense objects, and at their core, they harbor densities ranging from $10^{14}-10^{15}$ $g/cm^{3}$. Such high densities are suitable for testing the theory of strongly interacting matter where it is theorized that quark degrees of freedom are the stable ground state \cite{Witten,Olinto,Madsen,Rezolla,Annala,mallick1,mallick2,prasad1,prasad2}. However, observations from neutron stars are only from their surface, and we cannot directly probe their interiors. Therefore, it becomes necessary to model the neutron star not only from the core to the surface but also to understand its surrounding (namely the magnetosphere) from where the electromagnetic signals are being emitted and transported.

The first analytic theory of the electromagnetic wave emission assumed a rotating magnetized solid body as the star (not necessarily NS) \cite{deutsch}. A fast rotating magnetized star induces a strong electric field such that it can accelerate particles of the circumstellar medium to ultra-relativistic speeds, which emits electromagnetic radiation. The analytical result was interesting, which assumes the star is in a vacuum and emits monochromatic radiation. The modified Deutsch problem was subsequently solved, taking into account a co-rotating plasma up to the light cylinder \cite{kaburaki}. As the gravity at the surface of a neutron star is very high, it drags the frame around it as it rotates \cite{Morsink,Kamal}. 

The presence of plasma in the magnetosphere affects the pulsar radiation as then the radiation is multi-wavelength and leads to electromagnetically active neutron star model \cite{gold,pacini}. The power radiated for a perpendicular rotator in a vacuum was calculated, and even the physics of plasma was included in a multiplicative factor. Significant steps were then taken by Goldreich \& Julian \cite{julian} although for an aligned rotator. They were the first to argue that an empty magnetosphere cannot exist for a finite time as the induced electric field was strong enough to pull charged particles out of the star's surface and fill the magnetosphere. Also, it is likely to have a charge separation, and the polar caps were the regions emitting radio waves. Their calculations were further improved by Mestel \cite{mestel} for an oblique rotator, and the injection of particles at polar caps was subsequently done \cite{sturrock,jones}.\\
\indent The initial assumption of aligned rotating magnetic dipole losing energy via electromagnetic radiation in vacuum is unrealistic \cite{julian} as observation points towards broadband emission. To explain this and other observational aspects of pulsars like rotational braking, particle acceleration, and transport, one needs to understand the electrodynamics of pulsar magnetosphere.\\
\indent Usually, to understand the emission mechanism of pulsars, one starts with a vacuum magnetosphere empty of any plasma. For an aligned rotator, this will not emit any radiation; however, for a not aligned rotator, the emission is likely \cite{deutsch,michel}. However, their calculations were not totally general relativistic (GR) and assumed the star to be perfectly conducting. As already stated, vacuum magnetosphere is an unrealistic scenario to describe the pulsar mechanism of NS; therefore, subsequent calculation and recent simulation are done with plasma filled magnetosphere \cite{petri1,spitkovsky2006}. Force-free and ideal MHD conditions are still used; however, the simulation showed that such assumptions have their limitation.\\
\indent In most of these calculations, GR effects were not there implicitly. GR calculation involving static dipole magnetic field \cite{ginzburg,petterson}, multipole calculation \cite{anderson} and even for flat space-time has been done previously. The calculation for force-free monopole in Schwarzschild geometry also exists \cite{lyutikov}. Konno again revived the GR aligned rotator problem, and Kojima \cite{konno,Kojima}, and then more calculation involving electromagnetic field around a slowly rotating NS followed \cite{rezzolla2,zanotti,petri}. \\
\indent In the present calculation, we assume an oblique rotator in a vacuum with the force-free condition, perform a GR calculation, and show how the results change from the previous Newtonian calculations. The GR calculation for an oblique rotator is essential to understand the particle trajectory and their subsequent acceleration by electric field \cite{Kuzur,rupamoy}. In the next section (section II), we set up our problem and describe our formalism. In Section III, we present our results, and finally, in section IV, we summarize our results and draw conclusions.

\section{Formalism}

%\subsection{Defining The System}

\begin{figure}[h]
	
	\center{\includegraphics[width=60mm,scale=0.8]{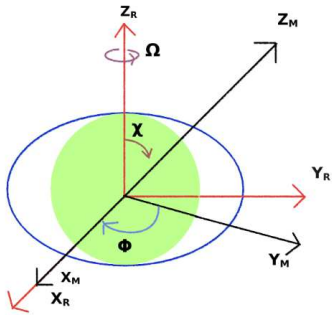}  }
	
	\caption{{\small{A schematic figure depicting the magnetic and rotation axes of a pulsar. Here $\chi$ is the misalignment angle with $\Omega$ being the frequency of rotation of the pulsar. The black-colored axes are the magnetic axes of the pulsar while the red colored ones are that of the rotation axes}}}
	
	\label{misalign}
	
\end{figure}

We start our calculation considering a misaligned pulsar as shown in fig \ref{misalign}. The angle $\chi$ is called the misalignment angle, the angle between the rotation and magnetic axis of the star. Coordinates of the magnetic field axis are coined as $(X_{M},Y_{M},Z_{M})$ and the rotational axes are denoted as $(X_{R},Y_{R},Z_{R})$.

In our model, we have made the following assumptions:

\begin{itemize}
	
	\item We consider a fast-spinning pulsar ($560$ Hz), which results in the star's deformation as shown in the figure. 
	
	\item The magnetic field of the pulsar is not strong enough to generate any deformation. 
	
\end{itemize}

%\section {Calculations For A Misaligned Rotating Star}

To find the magnetic field of a fast-spinning pulsar, we solve Maxwell's equation.

\begin{equation} \label{Maxwell}
\begin{aligned}
&\partial_{\mu} F^{\mu \nu} = \frac{4 \pi }{c} J^{\nu}, \\
\end{aligned}	
\end{equation}

where $F_{\mu \nu}= \partial_{\mu}A_{\nu} - \partial_{\nu}A_{\mu}$ is the electromagnetic tensor and $J^{\mu}$ is the 4-current. $A_{\mu}$ represents the magnetic vector potential.

To describe the structure of a rotating star, we assume the star's space-time to be defined by the Cook-Shapiro-Teukolsky metric \cite{Cook}. The space-time element for such a metric is defined as  \\

\begin{equation}
ds^{2} = -e^{\gamma + \rho} dt^{2} + e^{2 \alpha}(dr^{2} + r^{2} d\theta ^{2}) + e^{\gamma - \rho} r^{2} \sin ^{2} \theta (d \phi - \omega dt)^{2} 
\end{equation}

Here the metric potentials $\gamma$, $\rho$, $\alpha$ are functions of $r$ and $\theta$, and $ \omega $ (also a function of $r$ and $\theta$) is the frame dragging term. \\

To solve for the magnetic field we assume the 4-current to be of the form $J_{\mu}=(0,0,J_{\theta} \: \sin \chi,J_{\phi} \: \cos \chi)$, and similarly the magnetic vector potential is defined as $A_{\mu}=(0,0,A_{\theta} \: \sin \chi,A_{\phi} \: \cos \chi)$. The aligned rotator model is obtained with $\chi = 0$. The magnetic vector potential and the 4-current can be expanded in spherical harmonics as

\begin{align} \label{MisalignA}
& A_{\phi} = \sum_{l=1}^{\infty} a_{l}(r) \sin \theta \frac{d P_{l}(\cos \theta)}{d \theta} \nonumber  \\
& A_{\theta} = \sum_{l=1}^{\infty} b_{l}(r) \sin \theta \frac{d P_{l}(\cos \theta)}{d \theta} \nonumber \\
& J_{\phi} =\sum_{l=1}^{\infty} j_{a_{l}}(r) \: \sin \theta \: \frac{d P_{l} (\cos \theta)}{d \theta} \nonumber \\
& J_{\theta} =\sum_{l=1}^{\infty} j_{b_{l}}(r) \: \sin \theta \: \frac{d P_{l} (\cos \theta)}{d \theta} 
\end{align}

where, $P_{l}$ is the Legendre polynomial of degree $l$. Assuming the field to be dipolar ($l=1$) and substituting the values of the vector potential and 4-current in eq \eqref{Maxwell}, we obtain  

\begin{align} \label{Eq1}
& a_{1} \big[\omega \: \{1 + 3 \cos 2 \theta - \sin 2 \theta \: \frac{d \rho }{d \theta}\} + \sin 2 \theta \frac{d \omega}{d \theta} \big] \nonumber + \\ 
&r \sin ^{2} \theta \big[4\: e^{2 \alpha} \: \pi \: r \: {j_{a_{1}}} \: \omega + \omega \{r \: a_{1}'' + a_{1}' (2- r \frac{d \rho}{dr})\} + r \: a_{1} ' \frac{d \omega}{dr} \big] = 0	
\end{align}

\begin{align} \label{Eq2}
& a_{1} \big[2 e^{2 \rho} \: (-1 + \cot \theta \frac{d \rho}{d \theta})- r^{2} \omega \{ \omega (1 + 3 \cos 2 \theta - \sin 2 \theta \frac{d \rho}{d \theta}) + 4 \cos \theta \sin \theta \frac{d \omega}{d \theta}) \}] \nonumber \\ 
& r^{2} [e^{2 \rho} \{4 \: e^{2 \alpha} \pi {j_{a_{1}}} + a_{1}'' + a_{1}' \frac{d \rho}{dr}\} +\omega (r a_{1} '' + a_{1} ' (2 - r \frac{d \rho}{dr})) + 2\:r a_{1} ' \frac{d \omega}{dr} \big] = 0 \nonumber \\
\end{align} 

\begin{align}
3 \cot \theta - 2 \frac{d \alpha}{d \theta} + \frac{d \gamma}{d \theta} = 0
\end{align}

\begin{align} \label{Eq3}
4 e^{2 \alpha} \pi {j_{b_{1}}} + b_{1}'' + b_{1} ' (- 2 \frac{d \alpha}{dr} + \frac{d \gamma}{dr}) =0
\end{align}

Solving eq \eqref{Eq1} and eq \eqref{Eq2} we have

\begin{align} \label{Eq4}
& 2 a_{1} \: e^{2 \rho} (-1 + \cot \theta \frac{d \rho}{d\theta}) - r^{2} \omega \big[a_{1} \sin 2 \theta \frac{d \omega}{d \theta} + r^{2}\: a_{1} ' \sin ^{2} \theta \frac{d \omega}{dr}\big] + \nonumber \\ 
& r^{2} e^{2 \rho} \{ 4e^{2\alpha} \pi {j_{a_{1}}} + a_{1} '' + a_{1}' \frac{d \rho}{dr}\} =0
\end{align}

To solve for the magnetic field, we solve eq \eqref{Eq3} and eq \eqref{Eq4} with the initial condition given by 

\begin{align}
& {j}_{a_{1}} = c_{a_{0}} r^{2} (\rho_{0}(r) + p_{0}(r)) \cos \chi \\
& {j}_{b_{1}} = c_{b_{0}} r^{2} (\rho_{0}(r) + p_{0}(r)) \sin \chi \\ 
& a_{1} \approx (\alpha_{0}\: r^{2} + O(r^{4})) \cos \chi \\
&b_{1} \approx (\beta_{0}\: r^{2} + O(r^{4})) \sin \chi  
\end{align}
Where $c_{a_{0}}$ and $c_{b_{0}}$ are arbitrary constants, $\rho_{0}$ and $p_{0}$ are the central density and the central pressure of the star. $\alpha_{0}$ and $\beta_{0}$ are also constants which are fixed at the boundary.\\
\indent The magnetic field can be obtained by taking the curl of the magnetic vector potential in eq \eqref{MisalignA}. The magnetic field for the CST metric is obtained as:

\begin{align} \label{MagField}
& B_{r} = -\frac{e^{\alpha} \cos \chi}{r^{2} \sin \theta \sqrt{e^{4 \alpha + \gamma - \rho}}} \frac{\partial A_{\phi}}{\partial \theta} \nonumber \\
& B_{\theta} =  \frac{r e^{\alpha} \cos \chi }{r^{2} \sin \theta \sqrt{e^{4 \alpha + \gamma - \rho}}} \frac{\partial A_{\phi}}{\partial r} \nonumber \\
& B_{\phi} = - \frac{\sin \chi \sqrt{e^{\gamma - \rho}}}{r \sqrt{e^{ 4\alpha + \gamma - \rho}}} \frac{\partial A_{\theta}}{\partial r} \nonumber \\
\end{align}

Substituting the values of the magnetic vector potential in eq \eqref{MagField} we get the magnetic field components as: 

\begin{align} \label{MagField2}
& B_{r} =  \frac{2 a_{1}\: \cos^{2} \chi \: e^{\alpha} \: \cos \theta}{r^{2} \sqrt{e^{4 \alpha +\gamma - \rho}}}\nonumber \\
& B_{\theta}= -\frac{a'_{1} \sin \theta \: \cos^{2} \chi \: e^{\alpha}}{r \sqrt{e^{4 \alpha + \gamma - \rho}}} \nonumber \\ 	
& B_{\phi} = \frac{b'_{1} \: \sin^{2} \theta \:\sin^{2} \chi \: \sqrt{e^{\gamma - \rho}}}{r \sqrt{e^{4 \alpha + \gamma - \rho}}} \nonumber \\
\end{align}

%We will use these magnetic field values to calculate the star's electric field and determine the Poynting flux radiated finally. For our calculations, we use the RNS code \cite{Stergioulas}.

%\subsection{Spin Down Energy Loss}  

As the star rapidly rotates, it radiates energy along the magnetic poles and spins down with time.

Our basic assumption to study the power loss is

\begin{itemize}
	\item particles near the surface of the star are moving with a velocity given by 
	\begin{align}
	&v=(0,e^{\alpha}\: r\: \sin \chi \: W, \: e^{\frac{\gamma - \rho}{2}} \: r \sin \theta \:(\Omega + \cos \chi \: Y))
	\end{align}
	
	where W and Y are parameters (having dimensions of angular velocity) which are assumed to be $W = 0.7 \: \Omega$ and $Y = 0.7 \: \Omega$.
	
	\item there is no external force on the particle at the surface of the pulsar (force-free condition). Hence, the Lorentz force is given by 
	\begin{align}
	&\vec{F} = 0 = q [ \vec{E} + \vec{v} \times \vec{B} ]
	\end{align}
\end{itemize}

The electric field can be calculated from the known magnetic field by the equation.

\begin{align}
& \vec{E} = -[\vec{v} \times \vec{B}]
\end{align}

The components of the electric field are given by 

\begin{align}
&E_{r} = e^{- \alpha}((\cos \chi \: (\Omega + \cos \chi \: Y)(-a'_{1}\: \cos \chi \: \sin ^{2} \theta))+(\sin^{2}\chi \: W (b'_{1} \sin^{2} \theta \sin 
\chi))) \nonumber \\
&E_{\theta} = -\frac{e^{- \alpha }}{r} \: \cos \chi \:(\Omega + \cos \chi \: Y)(-a_{1} \: \cos \chi \: \sin 2\theta) \nonumber \\
&E_{\phi} = -\frac{e^{- \frac{\gamma - \rho}{2}}}{r \: \sin\theta} \sin \chi \: \cos \chi\: W \: (-a_{1} \: \cos \chi \: \sin 2 \theta) \label{E_comps}
\end{align}

The Poynting vector can be calculated from the equation.

\begin{align}
&\vec{S} = \frac{1}{\mu_{0}} (\vec{E} \times \vec{B})
\end{align}

where each component takes the form

\begin{align}
&S_{r}=\frac{1}{r^{2}}\Big[\sin \chi  \cos \chi e^{ \alpha - \gamma}  \left(a_{1} \sin 2\theta  \cos \chi \right) \nonumber \\
&(\csc ^2\theta  \: W \: \cos \chi  e^{2 \alpha +\rho } \left(a'_{1} \sin ^2\theta  \cos \chi \right)+e^{\gamma (r,\theta )} (Y \cos \chi +\Omega )  \left(b'_{1} \sin ^2 \theta \sin \chi \right)) \Big] \\ \nonumber \\
&S_{\theta}=\frac{1}{r^4}\sin \chi  r e^{-3 \alpha -\gamma }  \Big[r^2 e^{\gamma} (-b'_{1} \sin ^2\theta  \sin \chi )
(W \sin ^2\chi  (-b'_{1} \sin ^2\theta  \sin \chi )-\cos \chi  (Y \cos \chi +\Omega ) \nonumber \\
&(a'_{1} \sin ^2\theta  \cos \chi ))+\csc ^2\theta \:  W \: \cos ^2\chi  \: e^{2 \alpha +\rho } (-a_{1} \sin 2\theta  \cos \chi )^2\Big] \\ \nonumber \\
&S_{\phi}=\frac{\cos \chi }{r^3 \:\sin \theta  }\: e^{-2 \alpha - \frac{\gamma -\rho}{2}} \: \Big[\cos \chi  (Y \cos\chi +\Omega ) (r^2 (-a'_{1} \sin ^2\theta  \cos \chi )^2+(-a_{1} \sin 2\theta  \cos \chi )^2) \\ \nonumber
&-r^2 W \sin ^2\chi  (-b'_{1} \sin ^2\theta  \sin \chi) (-a'_{1} \sin ^2\theta  \cos \chi )\Big]
\end{align}

Therefore, finally, the energy loss is given by

\begin{align}\label{GR_Energy}
& \frac{dE}{dt} = - \int \vec{S} \cdot d\vec{A} \\
\implies
\mid &\frac{dE}{dt} \mid = \int L_{0} \sin^{2} \chi \big[ \sin ^{3} \chi -  \frac{\Omega \: \cos^{2}\chi \: a'_{1}}{W \: b'_{1}}\big] \: R^{2} \sin^{2} \theta \: d\theta \: d\phi
\end{align} 

where, 

\begin{align}
& L_{0} = \frac{e^{-3 \alpha} \: \sin^{4} \theta \: W \: b'^{2}_{1}}{r}
\end{align}

The energy loss for the Newtonian case is given by \cite{pacini}

\begin{align} \label{SpitkovskyEqn}
&\frac{dE}{dt} =\frac{2}{3} \frac{\mu ^{2} \Omega^{4}}{c^{3}} \sin^{2} \chi
\end{align}

Where $\mu$ is the magnetic moment.

\section{Results}

The magnetic field, electric field, and power loss of the rotating star are solved numerically by constructing the structure of the entire star. As the magnetic field is coupled to a background rotating star, the metric potentials for the rotating configuration have to be calculated numerically from the center to the surface of the star using the rotating neutron star (RNS) code \cite{Komatsu}. RNS utilizes the Hachisu self-consistent field method or HSCF method for the computational purpose. The pressure equation for such a system in differential form is given by
\begin{align}
dP-(\rho+P)\left[d\ln u^t+u^tu_\phi d\Omega\right]=0
\label{hyd}
\end{align}
where $u^tu_\phi$ is the differential rotation of the NS. Eq (\ref{hyd}) is solved by constructing sequences of stable rotating models and calculating $R_e$ and $R_p/R_e$, which is the equatorial radius and the axes ratio of the star, respectively. Specifying an equation of state (EoS) along with the initial central density and the star's rotation, one can model a rotating pulsar using this code. The model's outcome specifies the metric potentials, along with the total mass and radius of the star.

An hadronic matter (HM) EoS is used for solving the RNS code with BSR \cite{BSR} parameter set constructed using the relativistic mean-field (RMF) method. The EoS is consistent with the constraints imposed by the recent observations from NICER \cite{ch2_riley,ch2_miller} and LIGO \cite{ch2_abbott} experiments (equatorial radius for a $1.4\;M_{\odot}$ star, $12.71^{+1.14}_{-1.19}\: Km$, tidal love number $\Lambda_{1.4}\le580$ and compactness $0.156^{+0.008}_{-0.010}$ with maximum mass limit $\ge2.01\;M_{\odot}$).\\
\begin{figure}[h]
	\begin{minipage}[h]{0.4\textwidth}
		\includegraphics[scale=0.45]{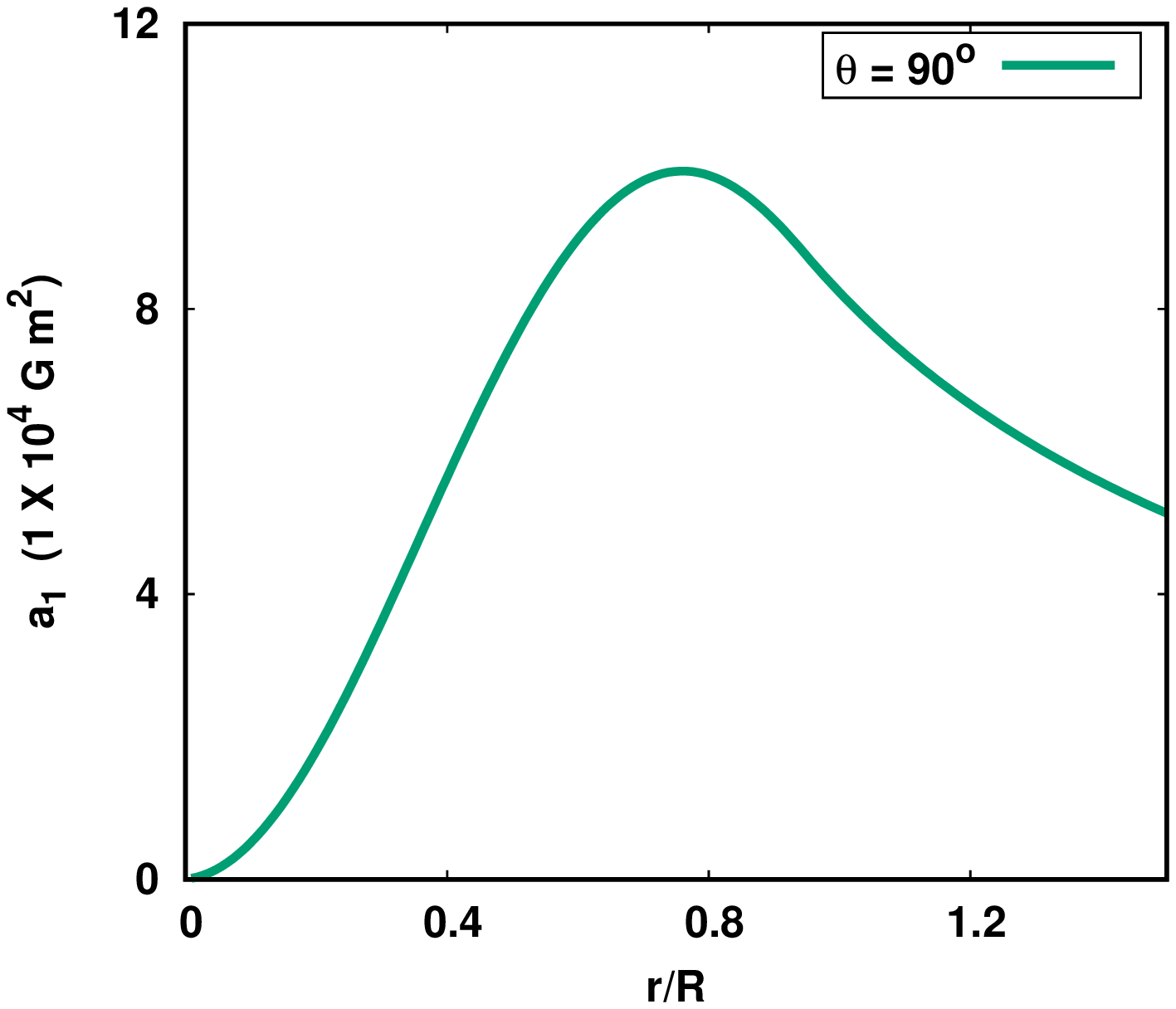}
	\end{minipage} \hspace{0.1\textwidth}
	\begin{minipage}[h]{0.4\textwidth}
		\includegraphics[scale=0.45]{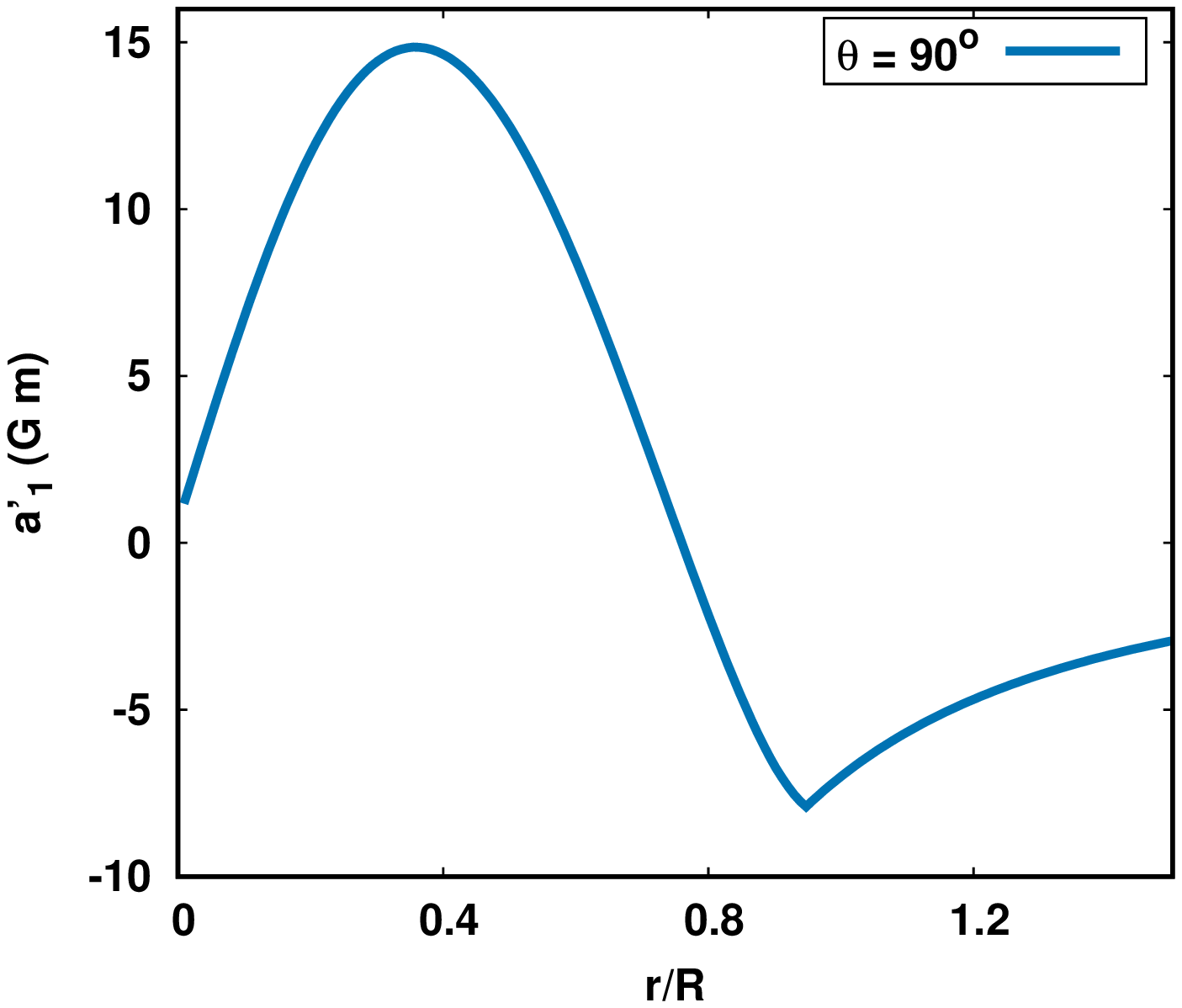}
	\end{minipage}
	\caption{\small {\bf (a):} The plot shows the nature of $a_{1}$ with varying radial distance. The normalized radius is the ratio of the radial distance to the radius of the star, for misalignment angle $0^{\circ}$ and $\theta = 90^{\circ}$. {\bf (b):} The plot shows the nature of $a'_{1}$ with varying radial distance. Where $a^{'}_{1}$ is the derivative of $a_{1}$ with respect to $r$. The normalized radius is the ratio of the radial distance to the radius of the star, for misalignment angle $0^{\circ}$ and $\theta = 90^{\circ}$. }
	\label{a1}
\end{figure}
%\subsection{Nature Of the Magnetic Field Lines}
\indent Using the RNS code with the BSR EoS parameter, we first solve the magnetic field. However, the magnetic field depends very strongly on $a_{1}$ and $a'_{1}$. The variation of them along the radial distance from the star center is shown in fig \ref{a1}.

\indent The variation is shown along the equatorial plane for an aligned rotator. Initially, $a_1$ is zero at the center and then gradually rises as we go radially outwards and attain a maximum inside the star. It then again goes down smoothly as we cross the star surface. Whereas $a'_{1}$ starts with a small non-zero value at the star center, attains a maximum at $0.3$R, and then goes down. It then attains a minimum at $0.9$R and smoothly rises as we cross the star surface.

\begin{figure}[h]
	\includegraphics[scale=0.45]{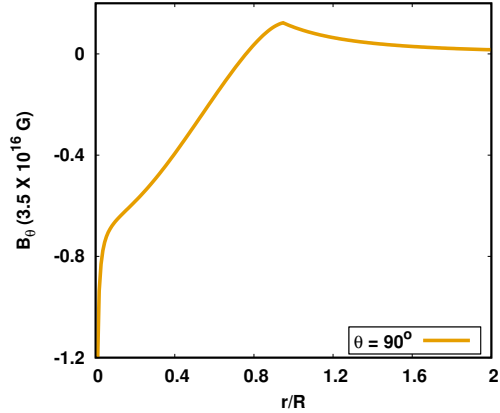}
	
	\caption{ The plot shows the nature of $B_{\theta}$ with varying radial distance. The normalized radius is the ratio of the radial distance to the radius of the star, for misalignment angle $0^{\circ}$ and $\theta = 90^{\circ}$. }
	\label{theta_90}
\end{figure}

Having studied $a_{1}$ and $a'_{1}$ we then plot the radial and polar component of the magnetic field in fig \ref{theta_90} for an aligned star at the equatorial plane ($\theta = 90^{\circ}$). Along the equatorial plane the $B_r$ component is zero as $B_{r} \propto \cos \theta$. $B_{\theta}$ is directly proportional to $-\frac{1}{r}$ and therefore initially falls off rapidly near the centre ; however, as we move away from the centre towards the surface the $a'_{1}$ term starts dominating ($B_{\theta} \propto - a'_{1}$) and similar to $a'_{1}$ there is switch in sign of the magnetic field at around $0.9$R.

\begin{figure}[h]
	\begin{minipage}[h]{0.4\textwidth}
		\includegraphics[scale=0.45]{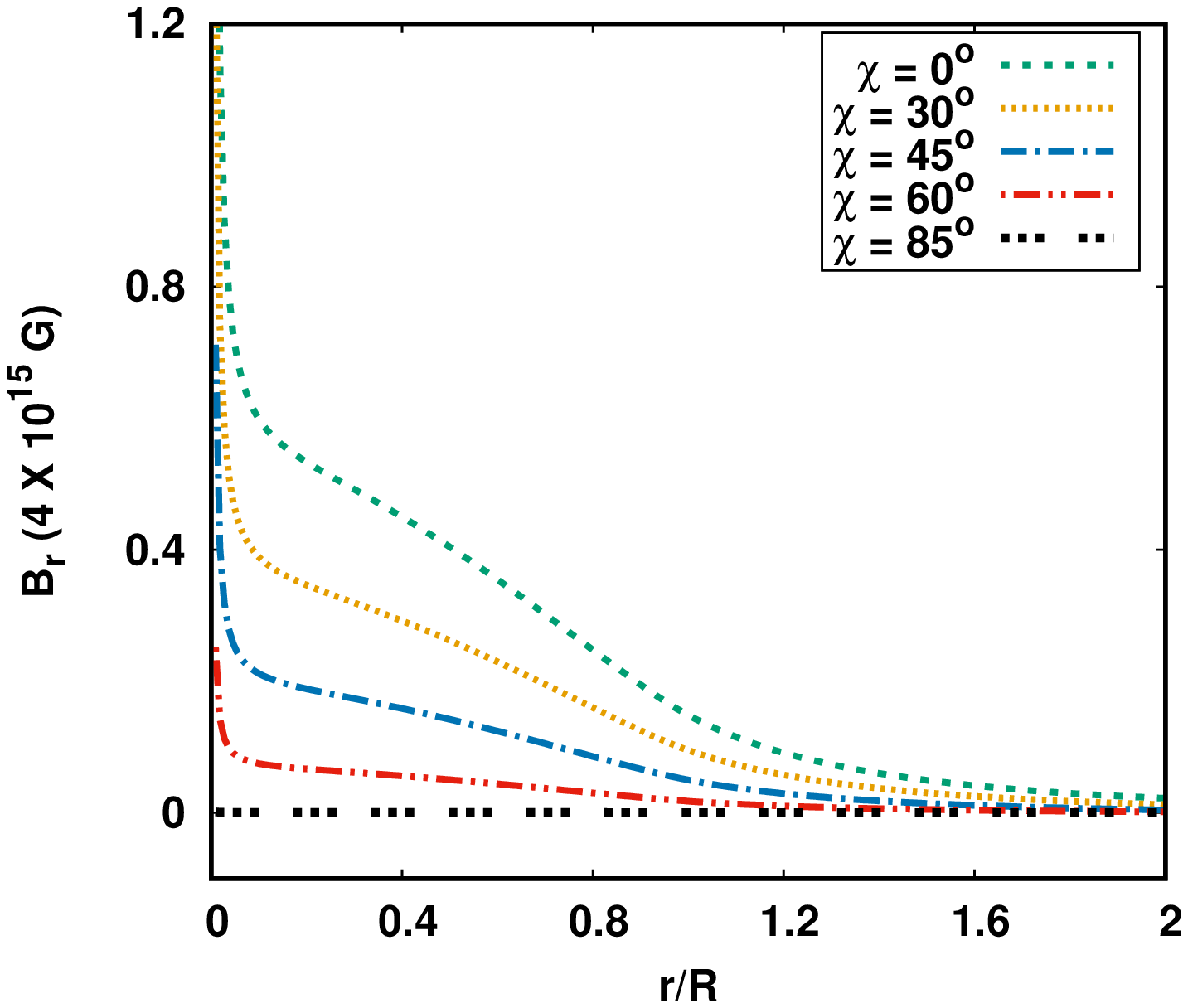}
	\end{minipage} \hspace{0.1\textwidth}
	\begin{minipage}[h]{0.4\textwidth}
		\includegraphics[scale=0.45]{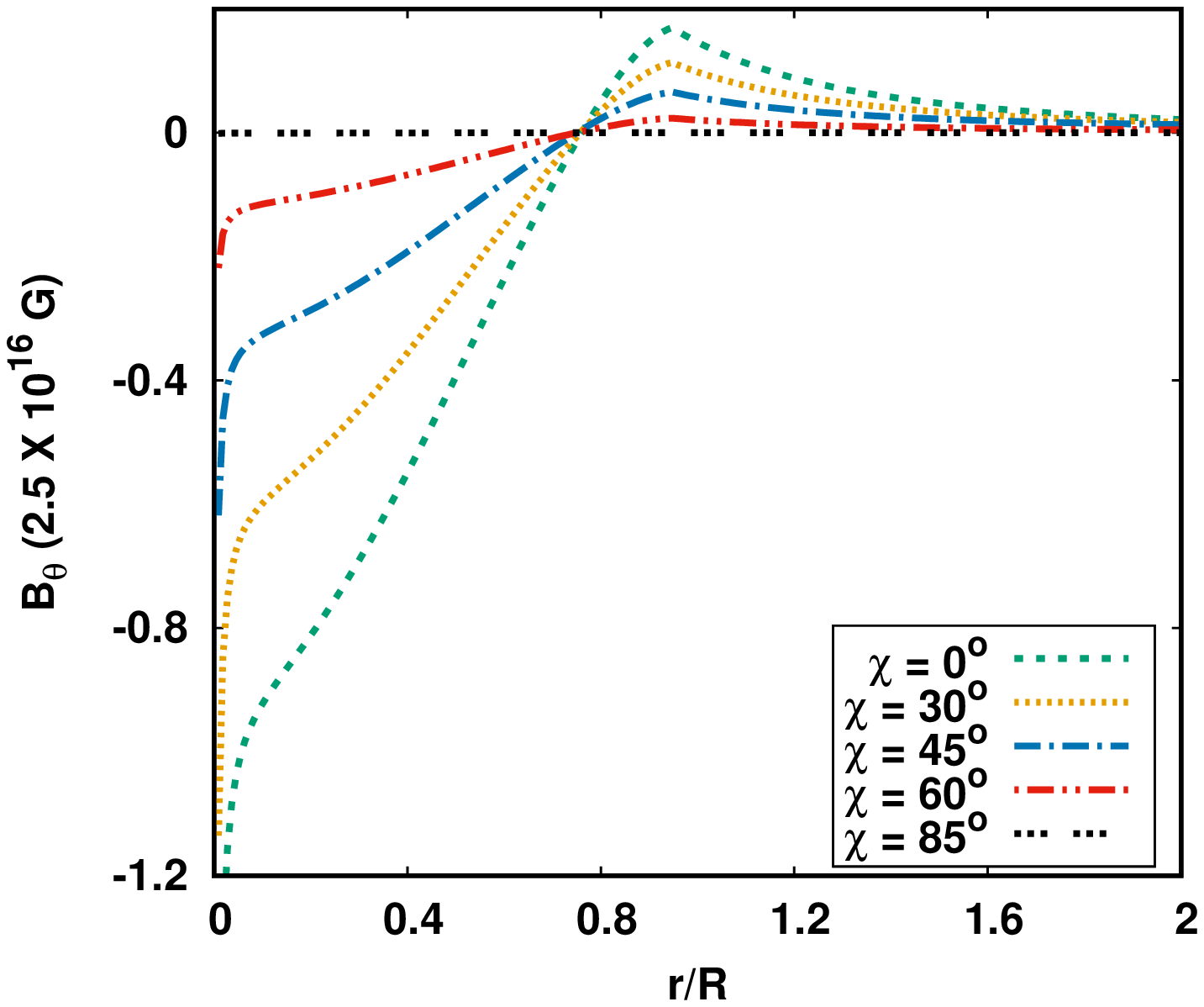}
	\end{minipage}
	\caption{\small {\bf (a):} The plot shows the nature of $B_{r}$ with varying radial distance. The normalized radius is the ratio of the radial distance to the radius of the star, for polar angle $\theta = 85^{\circ}$ for different values of $\chi$. {\bf (b):}The plot shows the nature of $B_{\theta}$ with varying radial distance. The normalized radius is the ratio of the radial distance to the radius of the star, for polar angle $\theta = 85^{\circ}$ for different values of $\chi$.}
	\label{theta_theta_85}
\end{figure}

\begin{figure}[h]
	\begin{minipage}[h]{0.4\textwidth}
		\includegraphics[scale=0.45]{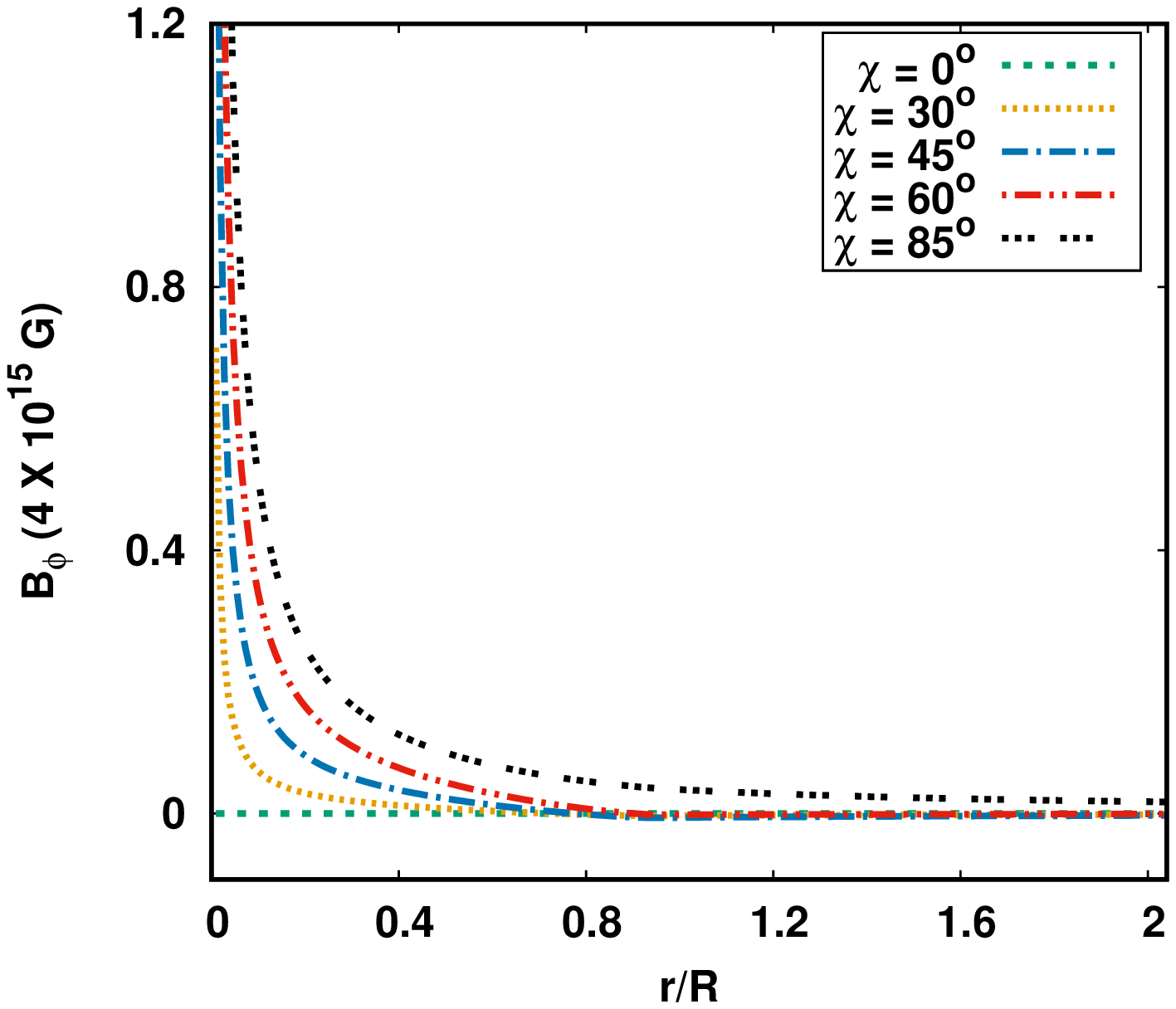}
	\end{minipage} \hspace{0.1\textwidth}
	\begin{minipage}[h]{0.4\textwidth}
		\includegraphics[scale=0.45]{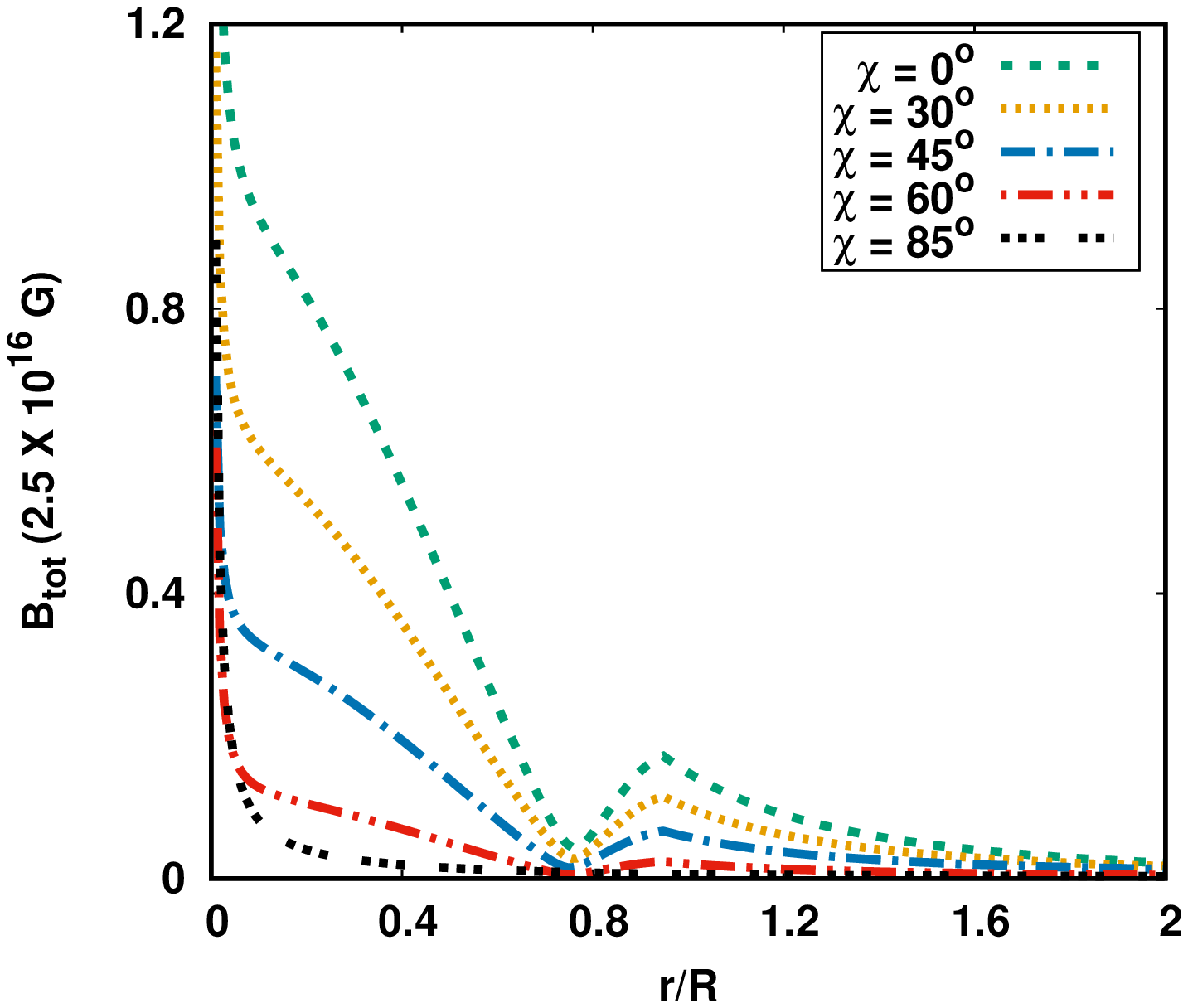}
	\end{minipage}
	\caption{\small {\bf (a):} The plot shows the nature of $B_{\phi}$ with varying radial distance. The normalized radius is the ratio of the radial distance to the radius of the star, for polar angle $\theta = 85^{\circ}$ for different values of $\chi$. {\bf (b):} The total magnetic field for the pulsar at different misalignment angle $\chi$ for a particular polar angle of $\theta = 85^{\circ}$.}
	\label{TotMag_theta_85}
\end{figure}

The results shown until now were for an aligned rotator; however, we need to have an oblique rotator to model a pulsar. The observed misalignment of pulsars is not the same and varies from a few degrees to a completely orthogonal rotator. Therefore, in fig \ref{theta_theta_85} we show the variation of $B_{r}$ and $B_{\theta}$ for different misalignment angles very near the equatorial plane ($\theta = 85^{\circ}$, not exactly equatorial plane as at equatorial plane $B_r$ is zero) . We observe that the magnetic field strength of the $B_r$ component reduces with an increase in $\chi$; however, beyond the star's surface, their values become very similar. The same nature is also observed in fig \ref{TotMag_theta_85}b for $B_{\theta}$ component, where the aligned pulsar shows maximum magnetic field strength. For an aligned rotator, the $B_{\phi}$ component is zero; however, for an oblique rotator, $B_{\phi}$ is finite and is shown in fig \ref{TotMag_theta_85}a. It falls off sharply near the center and then attains a comparatively small value near the star's surface.

%\begin{figure}[h]

%	\center{\includegraphics[width=70mm,scale=0.8]{TotalMag_theta_90.eps}}

%	\caption{{\small{Plot for the total magnetic field lines for $\theta = 85^{\circ}$. The y-axis shows the magnitude of the total magnetic field strength in Tesla. The normalized radius is plotted along the x-axis. }}}

%	\label{TotMag_theta_90}

%\end{figure}

Fig \ref{TotMag_theta_85}b shows the total magnetic field strength very close to the equatorial plane($\theta = 85^{\circ}$). There is a bump around the normalized radius of 0.7R. The bump reduces with an increase in the misalignment angle and eventually disappears for $\chi = 85^{\circ}$. For smaller misalignment angles, the dominant magnetic fields are $B_{r}$ and $B_{\theta}$ and the change in the sign for $B_{\theta}$ induces a bump in the total magnetic field; however, as $\chi \to 90^{\circ} $, $B_{\phi}$ (which is smooth) becomes dominant and hence the bump disappears.

\begin{figure}[h]
	\begin{minipage}[h]{0.4\textwidth}
		\includegraphics[scale=0.45]{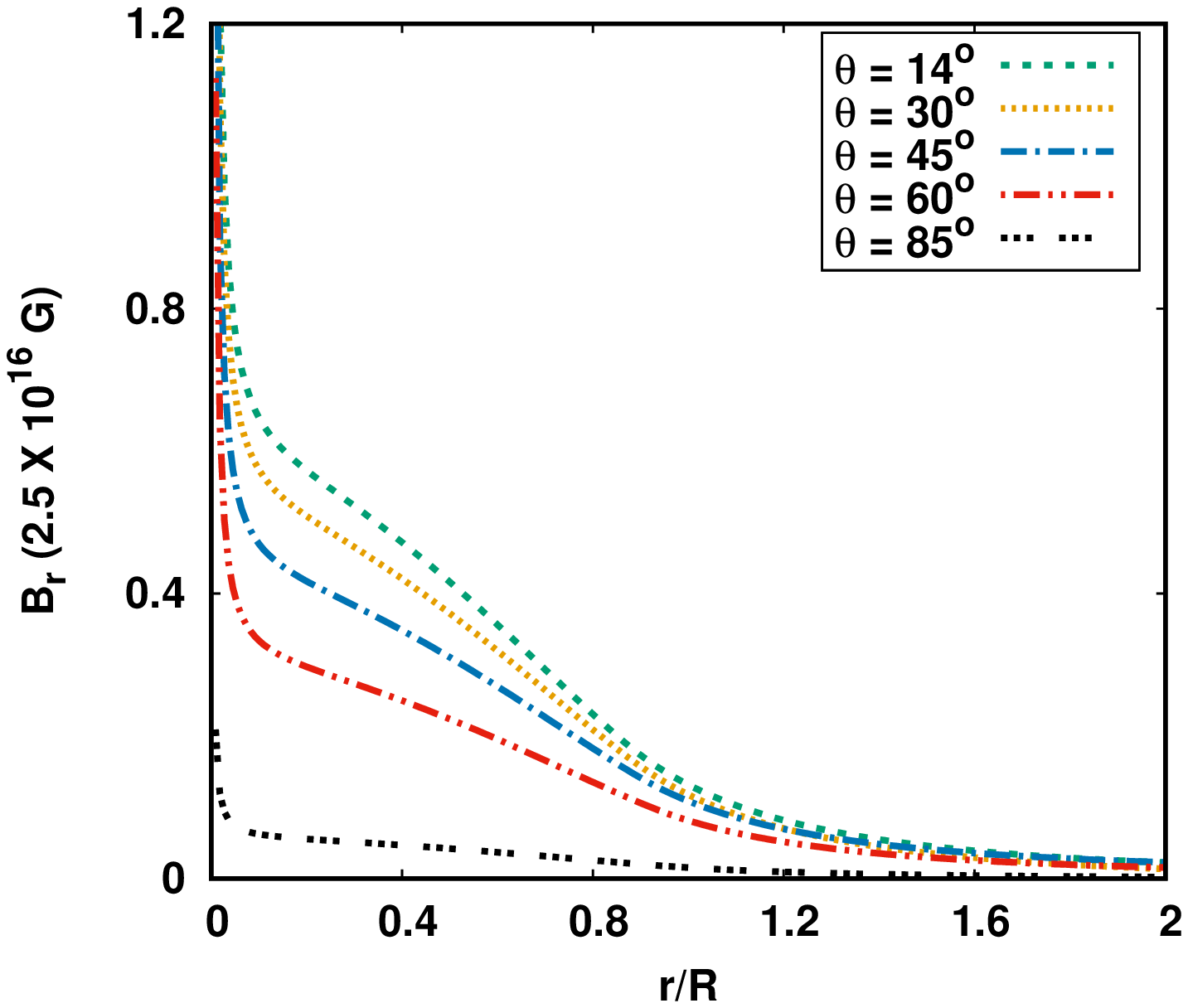}
	\end{minipage} \hspace{0.1\textwidth}
	\begin{minipage}[h]{0.4\textwidth}
		\includegraphics[scale=0.45]{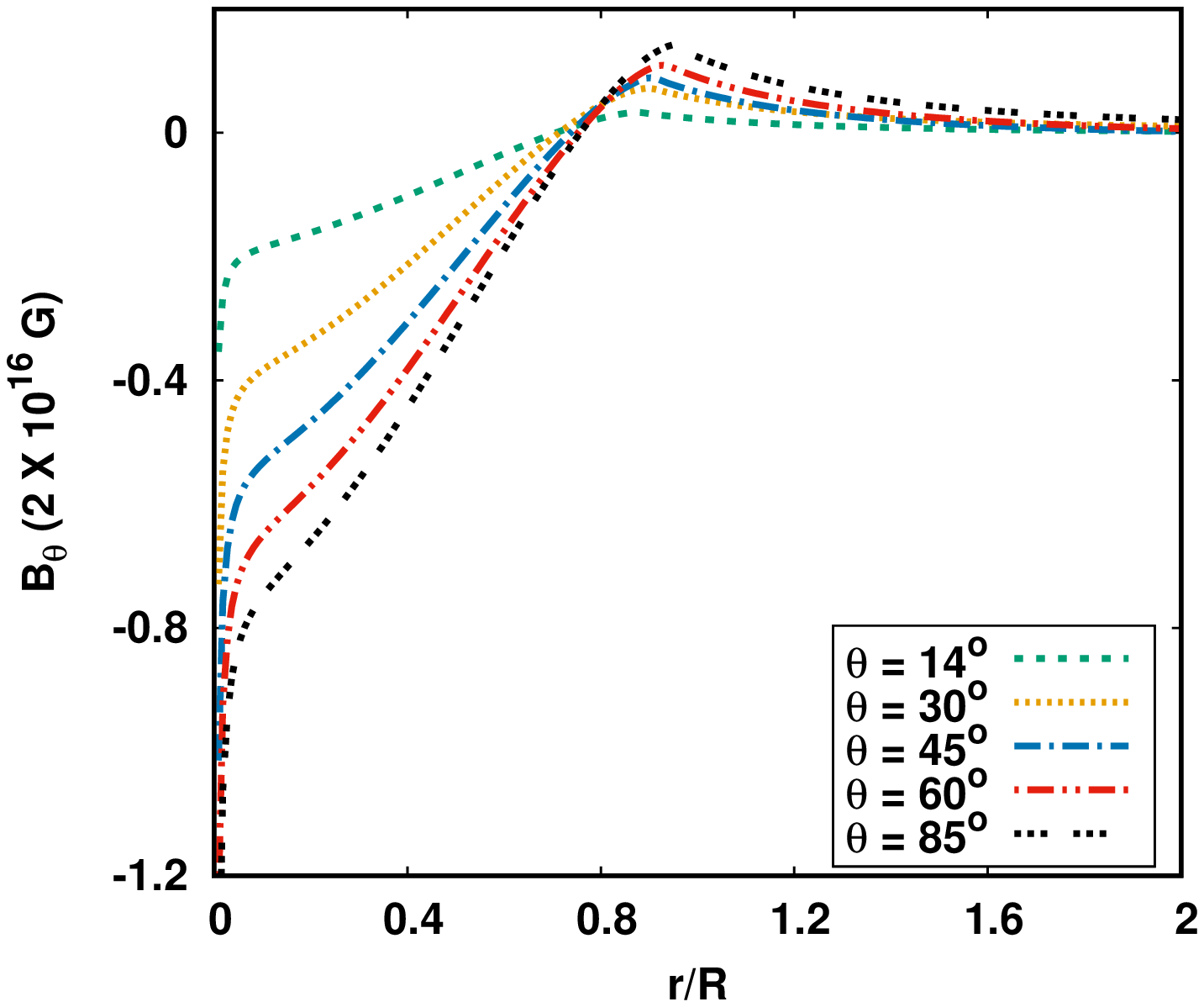}
	\end{minipage}
	\caption{\small {\bf (a):} This shows the plot for the r-th component of magnetic field at $\chi = 30^{\circ}$. The x-axis is the normalized radius. {\bf (b):}The plot shows the nature of $B_{\theta}$ with varying radial distance. The normalized radius is the ratio of the radial distance to the radius of the star, for misalignment angle $\chi = 30^{\circ}$ for different values of $\theta$.}
	\label{chi_chi_30}
\end{figure}

Fig \ref{chi_chi_30} shows the nature of the magnetic field lines for $r$-th and $\theta$-th components for different polar angles for an oblique rotator ($\chi = 30^{\circ}$). Since, $B_{r} \propto \cos \theta$, for lower values of $\theta$ the value of $B_{r}$ is higher. Similarly, $B_{\theta} \propto \sin \theta$ and hence with increase in $\theta$ the value of magnetic field also increases.

\begin{figure}[h]
	\begin{minipage}[h]{0.4\textwidth}
		\includegraphics[scale=0.45]{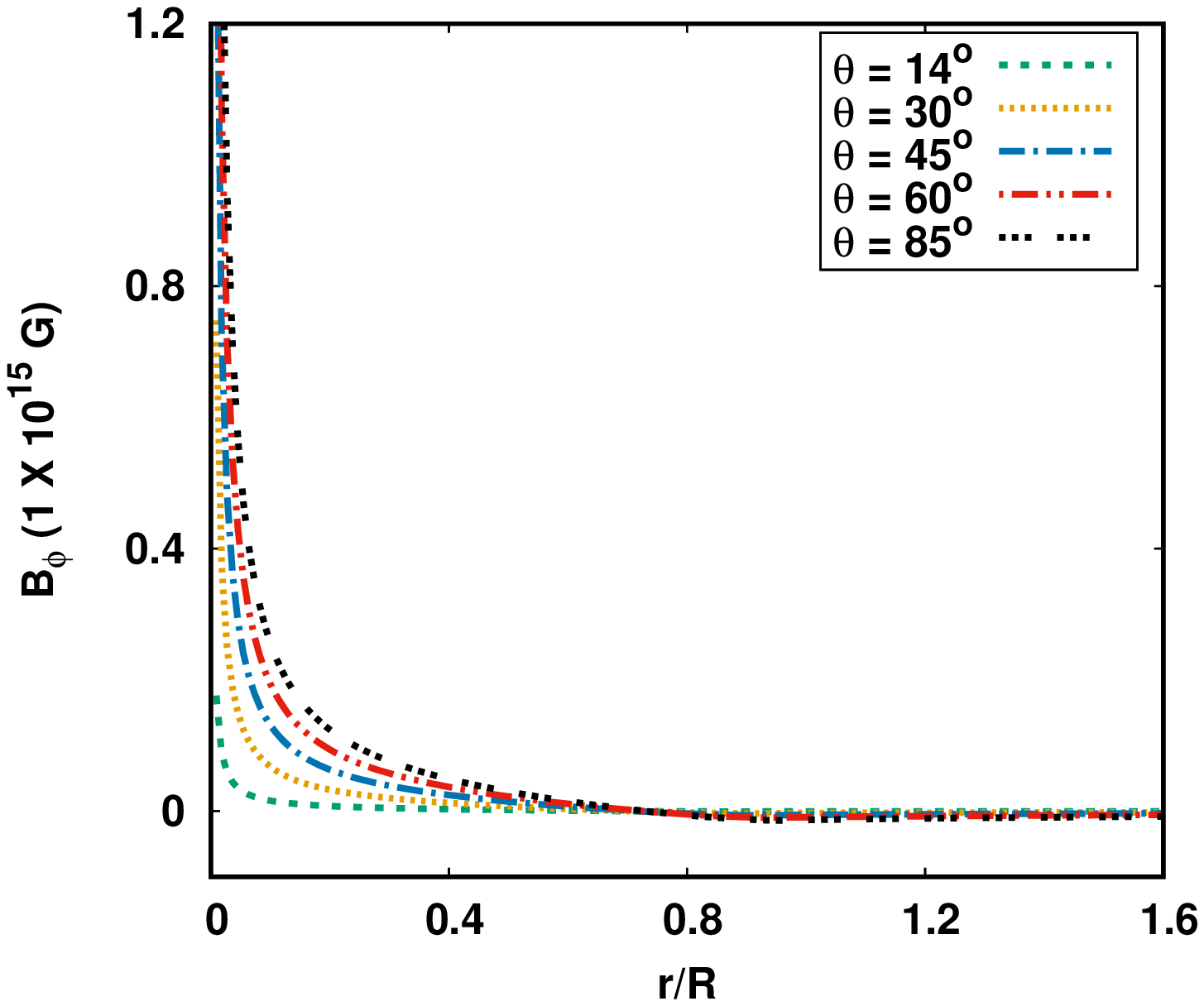}
	\end{minipage} \hspace{0.1\textwidth}
	\begin{minipage}[h]{0.4\textwidth}
		\includegraphics[scale=0.45]{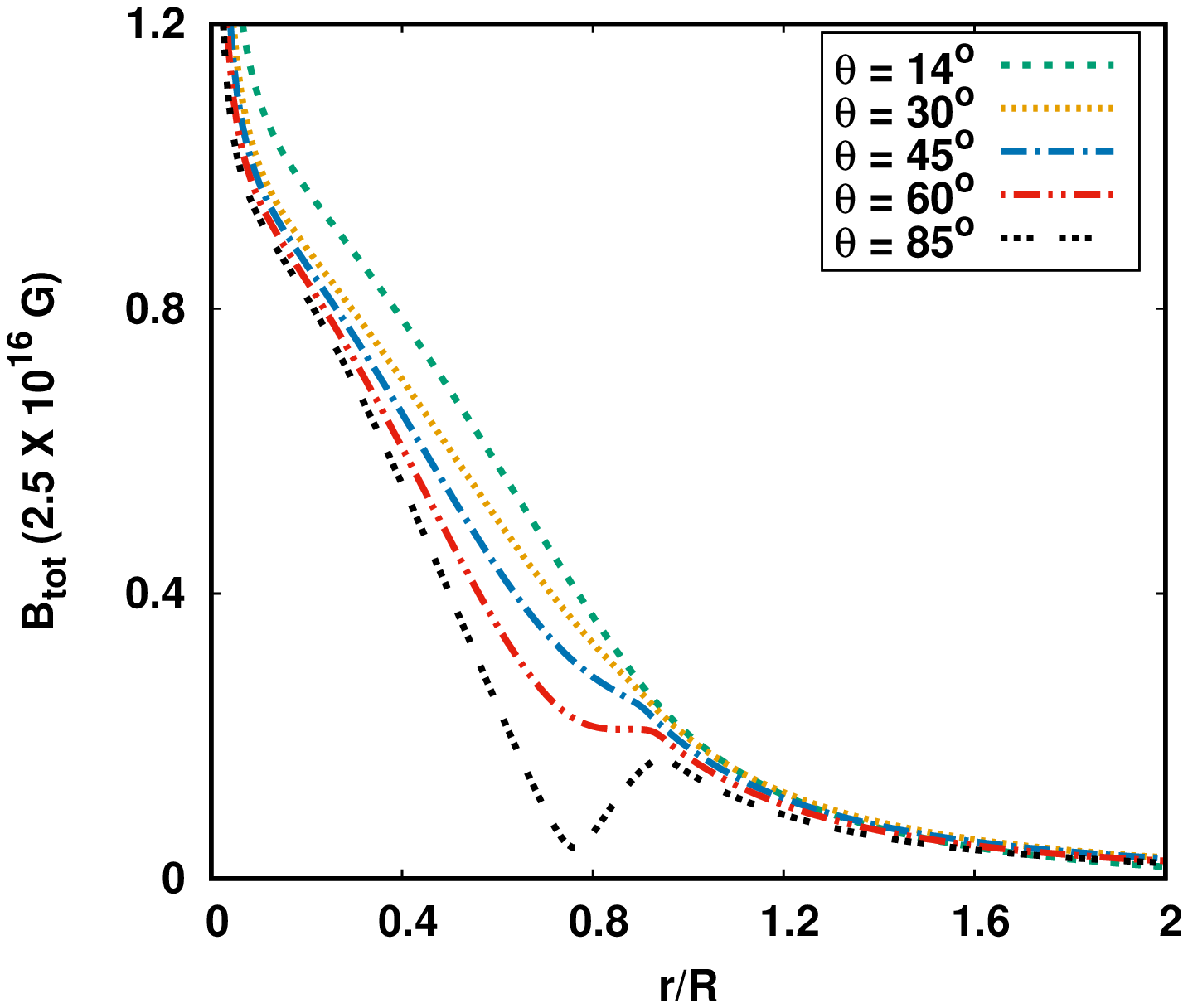}
	\end{minipage}
	\caption{\small {\bf (a):} Plot for $\phi$-component of magnetic field, $B_{\phi}$ at $\chi =30^{\circ}$ against the normalized radius. Different lines shows the the values of the magnetic field at different angles of $\theta$. {\bf (b):} The plot shows the nature of total magnetic field $B_{tot}$ with varying radial distance. The normalized radius is the ratio of the radial distance to the radius of the star, for misalignment angle $\chi = 30^{\circ}$ for different values of $\theta$.}
	\label{TotalMag_chi_30}
\end{figure}

Similar plot for $B_{\phi}$ is shown in fig \ref{TotalMag_chi_30}a, and we find that the magnetic field is strongest for $\theta = 85^{\circ}$, which is due to the fact that $B_{\phi} \propto \sin \theta$. The curves for the total magnetic field are shown in fig   \ref{TotalMag_chi_30}b, and it shows that the total magnetic field reduces rapidly for larger values of $\theta$. For $\theta = 85^{\circ}$ we observe that the value reduces and becomes zero at nearly $0.7$ R. Near the surface of the star, all the values of the magnetic field become nearly the same and eventually reduce to smaller values as we go further away from the surface of the star.

%\subsection{Results of power loss}
\begin{figure}[h]
	\center{\includegraphics[width=70mm,scale=1.0]{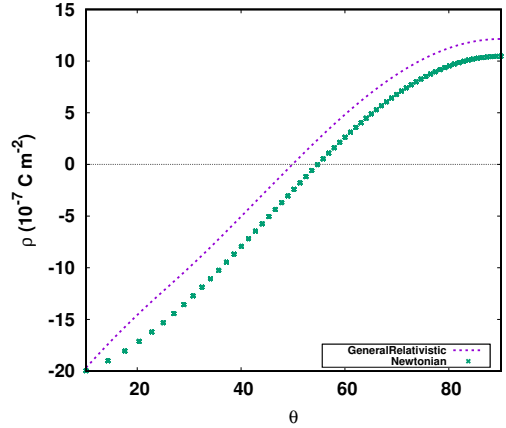}}
	\caption{{\small{The plot shows the comparison for the variation of Goldreich-Julian (GJ) density for the newtonian and the GR case. The y-axis shows the charge density and the x-axis shows the value of polar angle $\theta$. These two cases are for the aligned pulsar. }}}
	\label{Goldreich_Density}
\end{figure}

The rotation of the star results in a charge separation near the surface of the star \cite{julian}. This results in the accumulation of the negatively charged electrons near the pole, forming a dome of a shape \cite{Cerutti}. The positively charged particles, positrons accumulate around the equator of the star forming a torus. We compare the GJ density for the Newtonian calculation with our GR calculation for an aligned star. The Goldreich-Julian density is given by the expression,

\begin{equation}
\rho_{GJ} = - \frac{\vec{\Omega} .\vec{B} }{2\: \pi \:c}
\end{equation}
where $\vec{\Omega}$ and $\vec{B}$ are the angular velocity and the magnetic field vector, respectively. Speed of light is represented as c.
In our case, we obtain it to be

\begin{equation}
\rho_{GJ} = -\frac{\Omega \cos ^2\chi  \left(\sin ^2\theta \:  a'_{1}\: r\: e^{ \alpha}+2 a_{1} \cos ^2\theta \: e^{2 \alpha }\right)}{2 \: c \: \pi \: r^{2} \: \sqrt{ e^{4 \alpha +\gamma -\rho}}}
\end{equation}\label{Gold_GR}

Fig \ref{Goldreich_Density} shows the plot for the comparison of the Newtonian and GR case for an aligned star. For small $\theta$ (near the pole), the charge density is negative, indicating that the electrons are accumulated near the poles of the pulsar. As $\theta$ increases, the negative charge density decreases, and around $51^{\circ}$, it changes sign. At $51^{\circ}$ there is a charge-neutral point and beyond that positive charge accumulates. The nature of the plot is similar for both cases, but there is a difference in the values due to GR corrections. For the Newtonian case, the charge density is zero at $\theta \approx 55^{\circ}$, whereas, for the GR case, it is at $\theta \approx 51^{\circ}$.

\begin{figure}[h]
	
	\center{\includegraphics[width=70mm,scale=1.0]{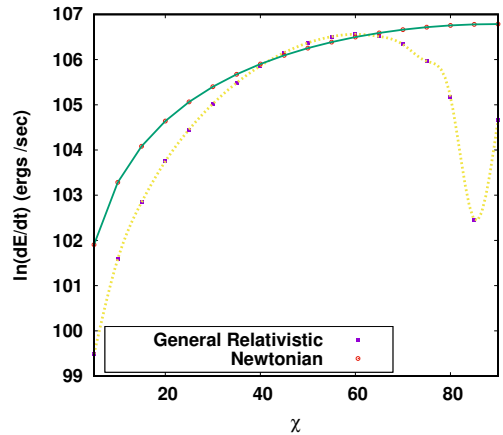}}
	
	\caption{{\small{The plot shows the variation of power loss for GR and Newtonian cases. The y-axis shows the logarithmic plot of the power loss in erg/sec, and the y axis shows the variation of misalignment angle $\chi$. }}}
	
	\label{PowerLoss}
	
\end{figure}

Once the magnetic field is known, it is trivial to calculate the electric field and also the power loss by the oblique rotator. The power loss for our calculation (GR) is shown in fig \ref{PowerLoss}, and for comparison, we also plot the power loss for the Newtonian calculation. The rate of power loss keeps on increasing with an increase in misalignment angle for the Newtonian case, whereas, for the GR case, initially, the power loss keeps increasing with an increase in $\chi$ till $\chi = 70^{\circ}$. If the misalignment angle becomes larger, the power loss reduces, and at $\chi = 85^{\circ}$, we observe a sharp dip in the rate of power loss. Beyond that, the power loss once again increases. It is interesting to note that for lower values of misalignment angle, the Newtonian value of power loss is greater than that of the GR case. At $\chi = 40^{\circ} $ the GR power loss rate becomes similar to that of the Newtonian and remains such till $\chi = 60^{\circ}$. Therefore, we can deduce that the spin-down of a GR oblique rotator is minimum if it is either an aligned rotator or an orthogonal rotator and the spin-down is fast in the in-between values. This nature of oblique rotator for GR calculation is completely different from Newtonian rotator, where the spin-down of a pulsar increases with an increase in misalignment angle and is maximum for an orthogonal rotator.

The nature of the dip in the power loss for the GR case can be explained from the Poynting vector, which is the product of magnetic and electric fields at the surface of the star.  To understand the power loss we plot the components of electric field $(E_{r},E_{\theta},E_{\phi})$ in the subsequent figures.

\begin{figure}[h]
	\begin{minipage}[h]{0.4\textwidth}
		\includegraphics[scale=0.5]{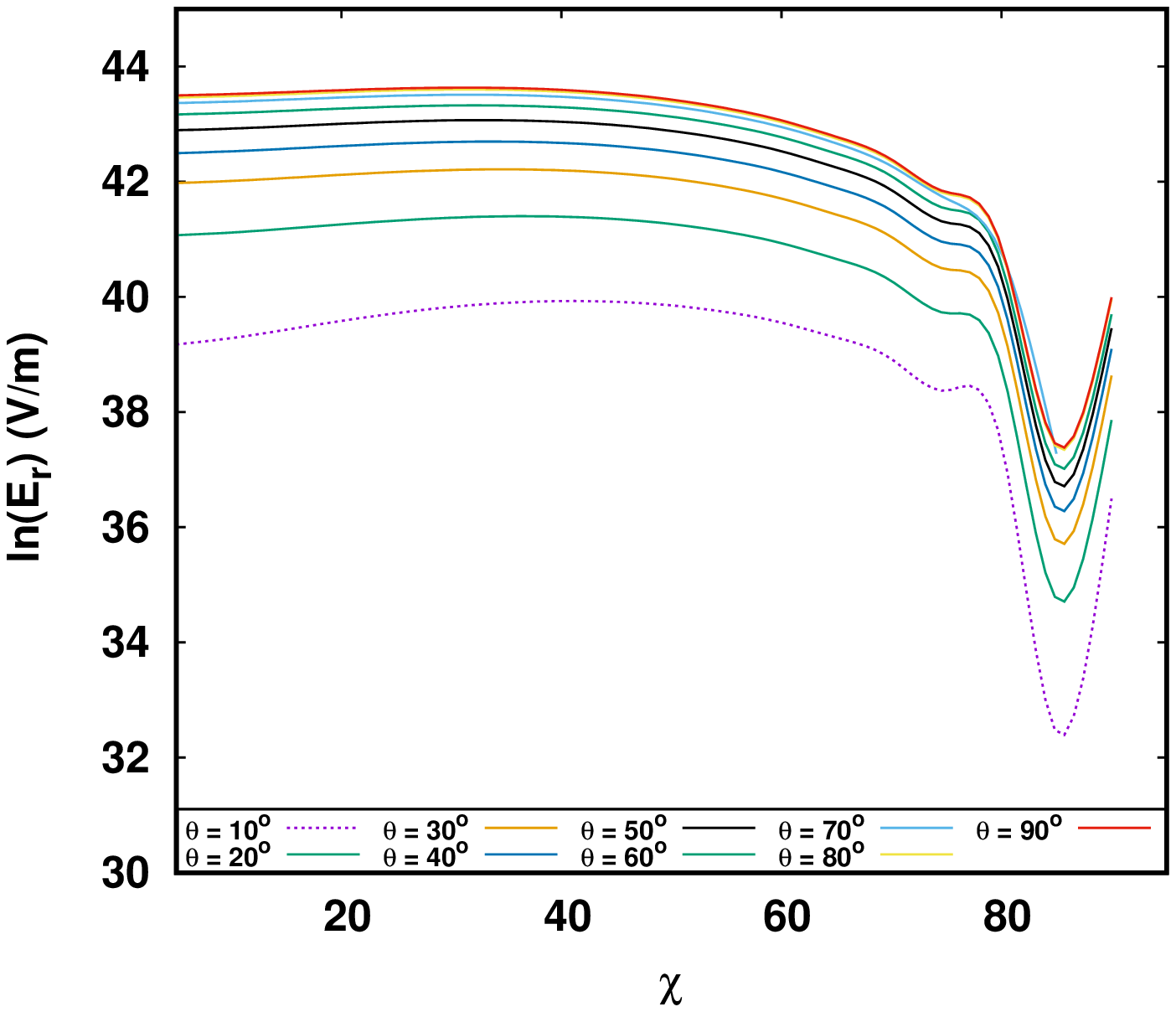}
	\end{minipage} \hspace{0.1\textwidth}
	\begin{minipage}[h]{0.4\textwidth}
		\includegraphics[scale=0.5]{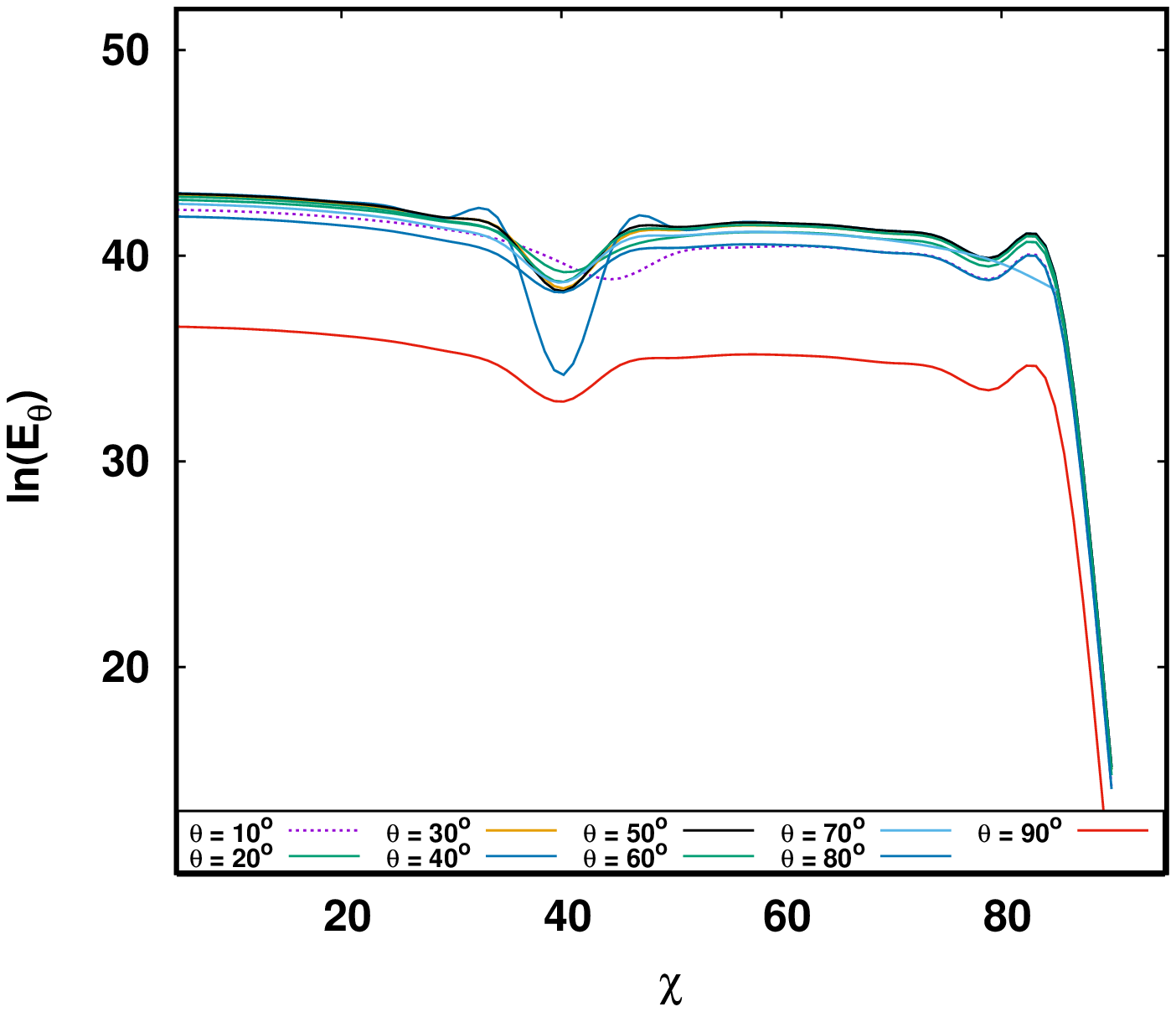}
	\end{minipage}
	\begin{minipage}[h]{0.4\textwidth}
		\includegraphics[scale=0.5]{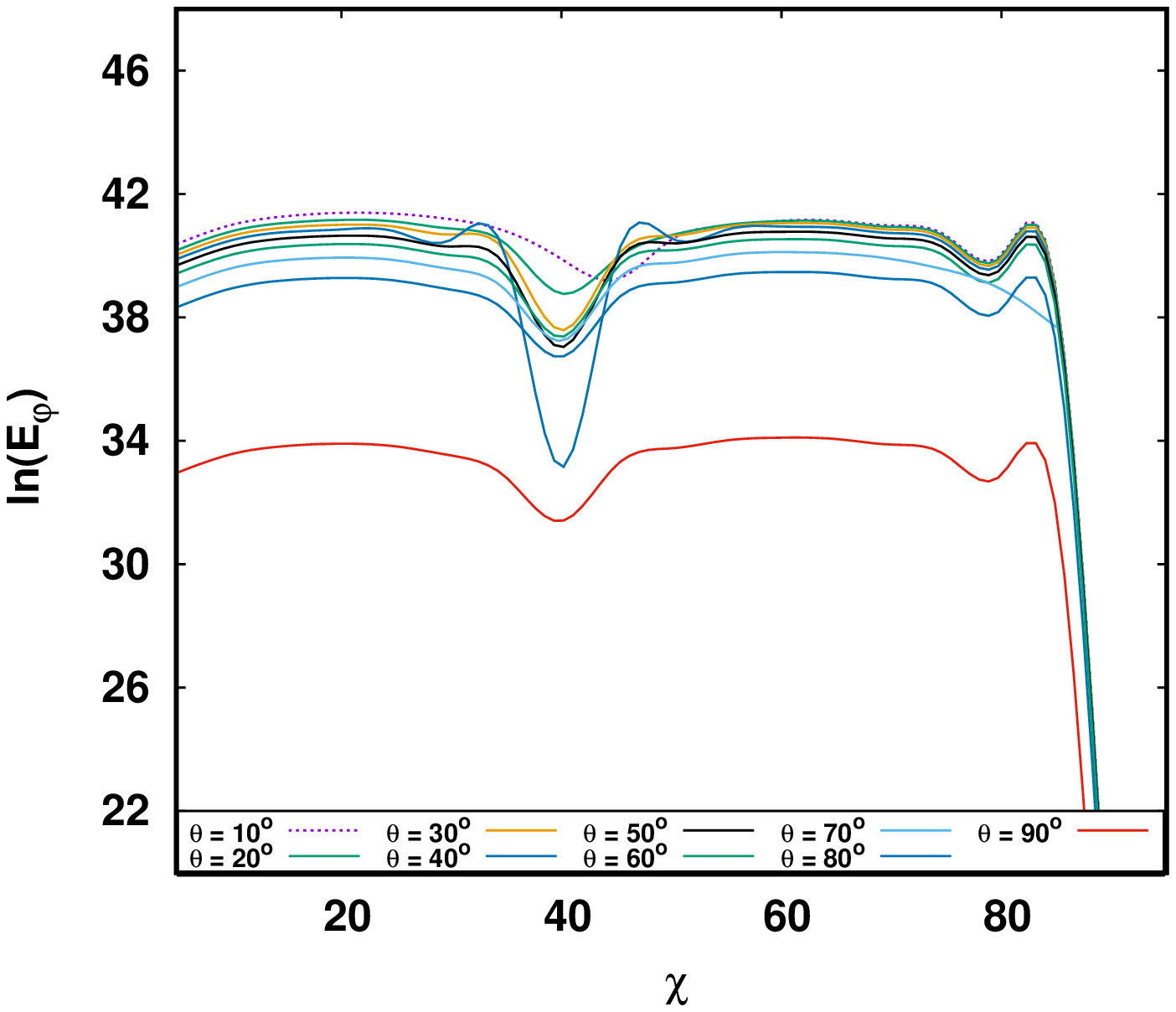}
	\end{minipage}
	\caption{\small {\bf (a):} Plot for $r$-component of electric field, against the misalignment angle $\chi$. Different lines shows the the values of the logarithmic value of r-th component of electric field at different angles of $\theta$ with increasing values of $\chi$.  {\bf (b):} The plot shows the nature of $\theta$-component of electric field $E_{\theta}$ with varying misalignment angle $\chi$. The y-axis shows the logarithmic value of the $\theta$-component at the surface of the star with misalignment angle $\chi$ along the x-axis for different values of $\theta$.  {\bf (c):} The plot describes the nature of variation of $\phi$-component of electric field. The y-axis shows the logarithmic value of $E_{\phi}$ at the surface of the star for different values of $\theta$ and the x-axis shows the variation in misalignment angle $\chi$.}
	\label{E_rtheta}
\end{figure}

%	\begin{figure}[h]

%\center{\includegraphics[width=75 mm,scale=1.0]{E_phi.eps}}

%	\caption{{\small{The plot describes the nature of variation of $\phi$-component of the electric field. The y-axis shows the logarithmic value of $E_{\phi}$ at the surface of the star for different values of $\theta$, and the x-axis shows the variation in misalignment angle $\chi$. }}}

%	\label{E_phi}

%\end{figure}

Fig \ref{E_rtheta} shows that all the components of the electric field tends to decrease as $\chi \to 90^{\circ}$, except for $E_{r}$ which shows a dip at around $\chi \sim 85^{\circ}$. From eqn (\ref{E_comps}) we see that as $\chi \to 90^{\circ}$ the $E_{r}$ component of the electric field starts to dominate and the remaining components of electric field $(E_{\theta}$ and $E_{\phi})$ tends to zero. This characteristic is also reflected in the power loss, which shows a dip near $\chi \sim 85^{\circ} $. Both the Newtonian and the GR power loss is proportional to $\sin^2 \chi$, which is why they initially rise with $\chi$. However, other factors contribute significantly to the GR power loss, and it differs from the Newtonian case.

\section{Summary and Conclusion}

In this work, we have performed a GR study of the oblique rotator, which is the model for the observed pulsars. Our oblique rotator is a rotationally deformed NS whose magnetic axis is inclined at an angle with the rotation axis, and it is losing rotational energy in the vacuum through the magnetic poles. We have modeled our rotationally deformed star with the CST metric and solved it numerically with the RNS code. As the star is misaligned, it is expected to have all the magnetic field components, and therefore we have chosen our 4-current and magnetic 4-vector potential accordingly. Assuming the field to be dipolar, we have solved Maxwell's equation in the asymmetric background space-time to obtain the magnetic fields.

The magnetic field induces an electric field with the assumption that the charged particles near the surface of a star have velocity components both in the polar and azimuthal direction (because of the oblique rotator). Also, in our calculation, we have assumed a force-free condition where the electric force balances the magnetic force on a particle. With the force-free assumption, we calculate the corresponding electric field and the Poynting vector to calculate the power loss by the star in the vacuum. To perform our calculation, we have used the BSR EoS, which describes the neutron star's matter.

We find that the magnetic field decreases with distance from the center of the star. Also, as the misalignment angle increases, the strength of the magnetic field decreases. The magnetic field shows a bump around $0.7$R which occurs due to the change in sign of the polar component of the magnetic field.

The magnetic field also decreases with an increase in polar angle and is minimum along the equatorial plane. The electric field remains almost constant initially as a function of misalignment angle, and its polar and azimuthal component goes to zero beyond $ \chi \sim 85^{\circ}$. However, the radial component has a minimum at $ \chi \sim 85^{\circ}$ but again increases as $\chi \to 90^{\circ}$.

The GJ charge density for an aligned rotator is qualitatively similar for both Newtonian and GR cases.

The power loss for the GR oblique rotator is significantly different from a Newtonian oblique rotator. As the power loss is directly proportional to $sin^2 \chi$ for the Newtonian case, it increases continuously with misalignment angle. However, for the GR case, it is proportional to $sin^2 \chi$ along with other factors, therefore although initially, it increases with misalignment angle it, however, shows a dip $ \chi \sim 85^{\circ}$ similar to the radial component of the electric field. The spin-down rate for the GR oblique rotator decreases if it is either an aligned rotator or an orthogonal rotator, whereas a Newtonian oblique rotator spin-down rate decreases if it is an aligned rotator.

We must mention that although we have performed a GR calculation for an oblique rotator, the power loss of the NS is in a vacuum. Our magnetosphere does not have any plasma or charged particles to account for any induced electromagnetic radiation. However, practically the electric field at the surface of the star is strong enough to extract charged particles from the star's surface and fill the magnetosphere. Our immediate future project is to perform such a simulation involving GR.

\section{Acknowledgments}

The authors are grateful to the Indian Institute of Science Education and Research Bhopal for providing all the research and infrastructure facilities.  DK would also like to thank the CSIR Government of India for its financial support. We would also like to thank Mr. Kamal Krishna Nath for his helpful discussions.

%\newpage

\end{document}